\documentclass[preprint,unsortedaddress]{revtex4-2}

\usepackage[pdftex]{graphicx}
\usepackage{multirow,tabularx}
\usepackage{amsmath}
\usepackage{amsfonts}
\usepackage{mathrsfs}
\usepackage{amssymb}
\usepackage[utf8]{inputenc}
\usepackage[usenames]{color}
\usepackage{siunitx}
\usepackage{bm}
\usepackage{hyperref}
\usepackage[ruled,vlined]{algorithm2e}
\usepackage{comment}
\usepackage[vlined]{algorithm2e}

\newcommand\thalf{{\tfrac12}}

\begin{document}
\title{Measure of distance and overlap between two arbitrary ellipses on a sphere}
\author{Andraž Gnidovec}
\affiliation{Department of Physics, Faculty of Mathematics and Physics, University of Ljubljana, Ljubljana, Slovenia}
\author{Anže Božič}
\affiliation{Department of Theoretical Physics, Jožef Stefan Institute, Ljubljana, Slovenia}
\author{Urška Jelerčič}
\affiliation{Department of Chemical Engineering, Ilse Kats Institute for Nanoscale Science and Technology,
Ben Gurion University of the Negev, Beer-Sheva, Israel}
\author{Simon Čopar}
\affiliation{Department of Physics, Faculty of Mathematics and Physics, University of Ljubljana, Ljubljana, Slovenia}
\date{\today}

\begin{abstract}
Various packing problems and simulations of hard and soft interacting particles, such as microscopic models of nematic liquid crystals, reduce to calculations of intersections and pair interactions between ellipsoids. When constrained to a spherical surface, curvature and compactness lead to nontrivial behavior that finds uses in physics, computer science and geometry. A well-known idealized isotropic example is the Tammes problem of finding optimal non-intersecting packings of equal hard disks. The anisotropic case of elliptic particles remains, on the other hand, comparatively unexplored. We develop an algorithm to detect collisions between ellipses constrained to the two-dimensional surface of a sphere based on a solution of an eigenvalue problem. We investigate and discuss topologically distinct ways two ellipses may touch or intersect on a sphere, and define a contact function that can be used for construction of short- and long-range pair potentials. 
\end{abstract}

\maketitle

\section{Introduction}

It comes as no surprise that packing of ellipses and ellipsoids is a very thoroughly researched topic that appears in many different fields of research, both in experimental realizations and in numerical models used to study them. Ellipsoids appear in Gay-Berne (anisotropic Lennard-Jones) models~\cite{GayJG_JChemPhys74_1981} of liquid crystals as a coarse-grained replacement for the full molecular structure~\cite{LuckhurstG_MolecularPhysics80_1993,ZannoniC_JMaterChem11_2001,AllenMP_MolecularPhysics117_2019}, in colloidal dispersions with an anisotropic dispersed phase~\cite{vanDillenT_MaterialsToday7_2004,RollerJ_SoftMatter16_2020,RollerJ_ProcNatlAcadSci118_2021}, and in granular and jammed matter~\cite{DonevA_PhDthesis,DonevA_Science303_2004,ManW_PhysRevLett94_2005,DonevA_PhysRevE75_2007}, where random and optimal packings are of particular interest~\cite{ChaikinPM_IndEngChemRes45_2006,JinW_PhysRevE95_2017}. All these examples are, however, Euclidean -- yet many experimental systems call for a confinement of particles to a curved surface, often that of a sphere. Recent examples include packings of rods~\cite{Smallenburg2016} and ellipsoids~\cite{XieZ_SoftMatter_2021}, spherocylinder simulations of nematics~\cite{BatesMA_JChemPhys128_2008}, and proteins adsorbed on vesicles~\cite{Frost2007,Frost2008}. This calls for an adaptation of ellipse-ellipse intersection algorithms for use on a spherical surface. Such an algorithm would also allow answering the question of optimal packing: while the well-researched Tammes problem~\cite{ClareBW_ProcRSocLondA405_1986,SaffEB_TheMathematicalIntelligencer19_1997} considers optimal packings of circles on a sphere, a generalization from circles to ellipses of arbitrary aspect ratios can provide us with the packing fraction for hard ellipses, which so far remains an open question. Furthermore, an algorithm which can be applied to ellipses of different sizes and aspect ratios opens up the possibility to consider polydisperse systems.

The bread-and-butter of computing ellipse-ellipse interactions lies in detecting collisions and overlaps in simulations of hard particles~\cite{MicheleCD_JComputPhys229_2010}, and, for long-range interactions, measuring the closest distance between them~\cite{EveraersR_PhysRevE67_2003}. One of the widely used and cited algorithms developed by Perram et al.~\cite{PerramJW_JComputPhys58_1985,PerramJW_PhysRevE54_1996} has been used, optimized, and adapted in numerous ways and for various applications -- in two dimensions (for ellipses)~\cite{ParamonovL_JChemPhys123_2005,ZhengX_PhysRevE75_2007}, three dimensions (for ellipsoids)~\cite{ZhengX_PhysRevE79_2009,GuevaraRodriguezFd_JChemPhys135_2011,ChoiMG_Symmetry12_2020}, and was even generalized to hyperellipsoids~\cite{GilitschenskiI}. However, all these algorithms are limited to Euclidean space and cannot be applied to the spherical case without modification.

In this work, we present a new algorithm that tackles the previously unsolved question of computing the distance and detecting overlap of ellipses confined to the two-dimensional surface of a sphere. Spherical confinement poses interesting challenges to the algorithm. Stretching is not a linear operation on a sphere, and two ellipses can interact in topologically different ways -- if they interact at all. These situations differ strongly from the Euclidean case. We explain the intricacies of the spherical ellipse-ellipse interaction with examples, discuss the performance of the numerical algorithm and conclude by showing a few packing solutions.

\section{Numerical algorithm}
\subsection{Problem formulation}

First, we must define what constitutes an ellipse on the surface of a sphere. We adopt the conventional definition of an ellipse as the set of points with a constant sum of distances to the foci. To generalize it to a sphere, we require a constant sum of geodesic distances (great circle distances) to the foci. This definition is satisfied by an intersection of the unit sphere and an elliptical cylinder given in the form of a degenerate positive semidefinite quadratic form $a(\bm{r})$:
\begin{equation}
  a(\bm{r})=\bm{r}A\bm{r};\quad A=T\operatorname{diag}(0,\Box,\Box)T^{\rm T},
\end{equation}
where $T$ is a rotation matrix that will not be explicitly needed, as we assume from now on that $A$ is a given quantity which can be computed from any representation of the ellipses, such as from center vectors and major semi-axis direction or from Euler angles. Looking for an intersection of two arbitrary spherical ellipses is therefore equivalent to looking for an intersection of two quadratic forms and the unit sphere:
\begin{eqnarray}
a(\bm{r})&=&\bm{r}A\bm{r}=1,\\
b(\bm{r})&=&\bm{r}B\bm{r}=1,\\
\|\bm{r}\|&=&1.
\end{eqnarray}
However, the quadratic forms $a$ and $b$ are invariant to inversion. Both $a=1$ and $b=1$ give \emph{a pair} of antipodal ellipses when intersected with the unit sphere. This poses an additional challenge for the collision detection algorithm, as we must specify which ellipse is the correct one and which collisions to ignore. The correct ellipses can be specified by vectors $\bm{r}_A$ and $\bm{r}_B$ corresponding to the centers of the ellipses -- \emph{signed} eigenvectors corresponding to the zero eigenvalue of the quadratic forms $a$ and $b$. The dot product between the ellipse center and any point on the ellipse is positive for the correct ellipse and negative for the antipode.

Unlike in the Euclidean case, scaling the semiaxes of the quadratic form has an important effect on the topology of its intersection with the unit sphere. When the semiaxes are small compared to the radius of the sphere, the ellipses are similar to Euclidean ellipses. If the semiaxes are scaled to be comparable to the sphere radius, the apexes become sharper and converge to a ``lemon wedge'' shape in the limit where the large semiaxis of the quadratic form matches the sphere radius. In this configuration, the antipodes touch at two ``poles'', forming two intersecting great circles. Beyond this size, the intersection with the sphere splits again into a new pair of ellipses, but now their centers are directed along the shorter of the quadratic form semiaxes. At this crossover, the former antipodal pair recombines, and no longer correspond to elliptical particles centered at $\bm{r}_{A,B}$. These cases with \emph{inverted} ellipses will play a role in our theoretical analysis, but have no physical significance.

The goal of our algorithm is to detect when two ellipses are tangent or overlapping by defining a contact function and to obtain the contact point $\bm{v}$. If forces at the contact point are required, the direction of the force should be along the normal to the ellipse, which is given by the gradient of the quadratic form (magnitudes can be normalized -- here we halve the expression to simplify notation):
\begin{equation}
  \bm{n}=\thalf\nabla_\perp \bm{r}A\bm{r}\big|_{\bm{r}=\bm{v}} =
  A\bm{r}-\bm{r}(\bm{r}A\bm{r})\big|_{\bm{r}=\bm{v}}=(A-I)\bm{v}.
\end{equation}
From the force and the intersection point, we can also compute torques acting on the ellipse, which is useful for molecular dynamics simulations.

\subsection{Solving for ellipse contacts}

\begin{figure}
  \centering 
  \includegraphics[width=\linewidth]{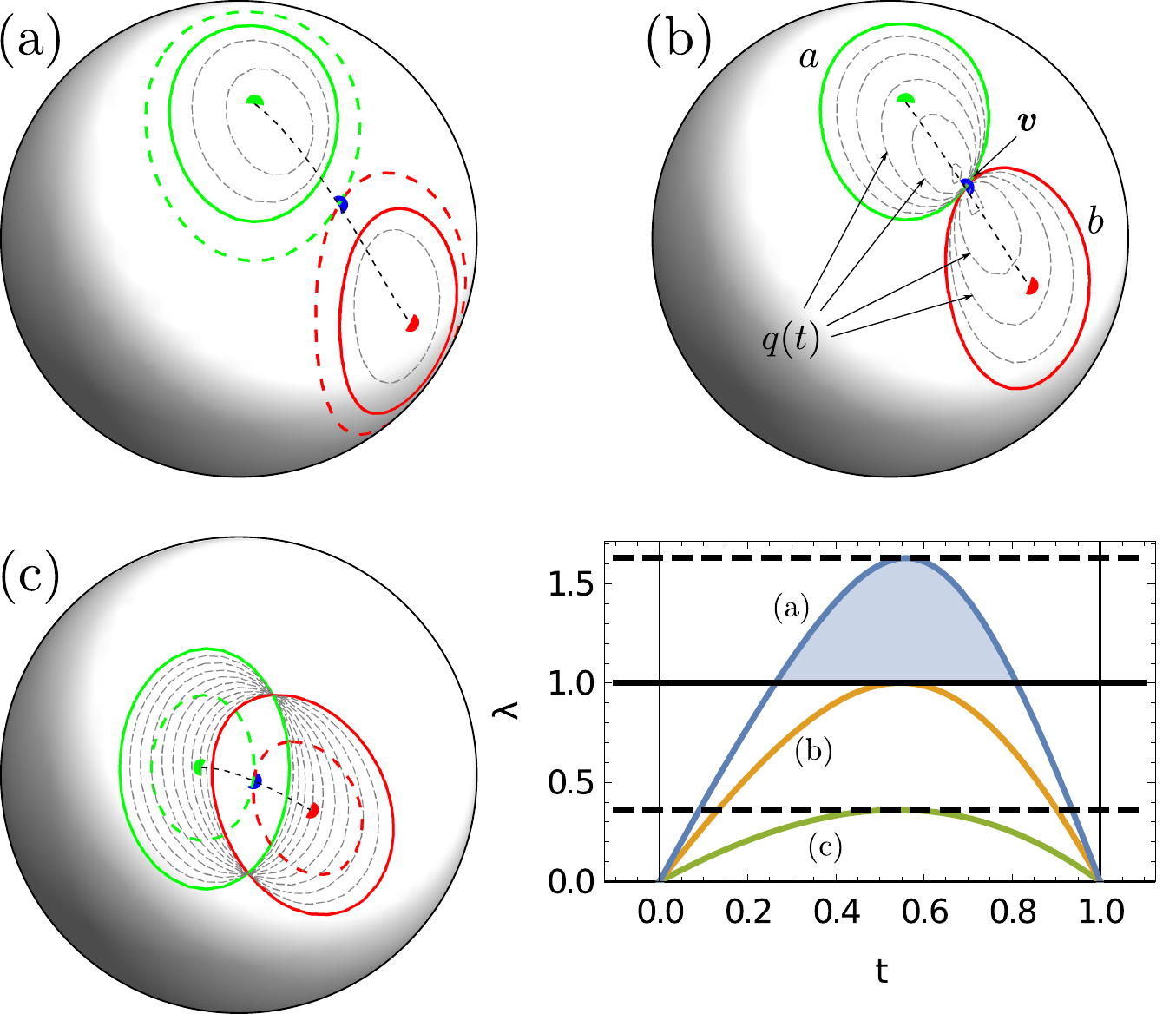}
\caption{Cases of {\bf (a)} disjoint, {\bf (b)} touching, and {\bf (c)} overlapping ellipses on a sphere, with ellipses stretched to achieve tangency shown by dashed colored lines. Level sets of $q(t)=1$ [Eq.~\eqref{eq:qt}] at different $t$ are shown by dashed gray lines. Black dashed lines represent the lines described by the eigenvector corresponding to the smallest eigenvalue of $Q(t)$ when $t$ runs from $0$ to $1$, and $\bm{v}$ marks the intersection point found when the eigenvalue is maximized. {\bf (d)} Smallest eigenvalue with respect to $t$ for the three cases in panels (a)--(c), showing that the smallest eigenvalue exceeds $1$ when the ellipses are disjoint.}
  \label{fig:stretch}
\end{figure}

Following the same steps as Perram and  Wertheim~\cite{PerramJW_JComputPhys58_1985,PerramJW_PhysRevE54_1996,DonevA_PhysRevE75_2007}, we define a linear interpolation of the quadratic forms with the parameter $t\in [0,1]$:
\begin{equation}
  q(t)=\bm{r}Q(t)\bm{r},\quad Q(t)= A (1-t) + B t
  \label{eq:qt}
\end{equation}
The value of a quadratic form $a$ constrained to the surface of the sphere is $a<1$ inside the ellipse and $a>1$ outside the ellipse, with the ellipse being the level set at $a=1$. We know that the level set of $q(0)=1$ on the unit sphere coincides with the first ellipse ($a=1$) and the level set of $q(1)=1$ with the second ellipse ($b=1$). Moreover, the value of the quadratic form $q(t)$ is always greater than $1$ in the part of the sphere that is outside both ellipses, as it is an interpolation of two values greater than $1$. Therefore, if the ellipses do not overlap, the space of allowed level set locations is discontinuous and the level set cannot evolve continuously from $t=0$ to $t=1$ and must thus be empty for some $t$. Conversely, if the ellipses intersect, the intersection points of $a=1$ and $b=1$ are a part of the level set for each $t$, thus ensuring that the level set is non-empty for each $t$. Examples of disjoint, touching, and overlapping ellipses and the level sets of $q(t)$ are shown in panels (a) to (c) of Fig.~\ref{fig:stretch}.

The unconstrained three-dimensional level set $q(t)=1$ represents a generic ellipsoid (or possibly a degenerate elliptical cylinder when one eigenvalue of $Q(t)$ is zero), and the eigenvalues of $Q(t)$ correspond to inverse squares of its semiaxes. This ellipsoid is thus completely contained within the unit sphere if \emph{all} its eigenvalues are greater than $1$ and intersects the unit sphere if this is not the case. It follows that level sets $a=1$ and $b=1$ intersect on the unit sphere if and only if the smallest eigenvalue of $Q(t)$ never exceeds $1$ on the interval $t\in[0,1]$ (see Fig.~\ref{fig:stretch}d).

We define the smallest eigenvalue and the corresponding eigenvector as
\begin{equation}
  \lambda_1(t)=\operatorname{eigmin}Q(t),
\end{equation}
then find the extremum $\Lambda_1$ of this eigenvalue and the corresponding eigenvector $\bm{v}_1$,
\begin{equation}
  \Lambda_1 = \operatorname{max} \lambda_1(t), \quad Q \bm{v}_1 = \Lambda_1 \bm{v}_1.
\end{equation}
Positive definiteness ensures there are always three nonnegative real eigenvalues, corresponding to the \emph{casus irreducibilis} of the cubic equation, which is solvable in closed form through trigonometry. To find the maximum $\Lambda_1$, any one-dimensional maximization algorithm can be used, such as the golden section search. We can rely on this function being anticonvex with a single maximum, which ensures reliable and fast convergence.

The value of $\Lambda_1$ has a clear geometric meaning: If we observe the intersection with a sphere $r^2=1/\Lambda_1$ instead of the unit sphere, the ellipses $a=1$ and $b=1$ touch at a single point of tangency, given by the appropriately scaled eigenvector $\bm{v}_1$. Scaling the system back to the unit sphere by a factor of $\sqrt{\Lambda_1}$, we see that $\Lambda_1$ is the factor by which the orthogonal projected area of both ellipses must be grown to make them tangent (scaling the semiaxes by $\sqrt{\Lambda_1}$). Values of $\Lambda_1>1$ signify non-overlapping ellipses which become tangent when grown, and values of $\Lambda_1<1$ overlapping ellipses which become tangent when shrunk. This property makes $\Lambda_1$ an appropriate choice for a contact function, with the same meaning it has in the Euclidean case (see the work of Perram and Wertheim~\cite{PerramJW_JComputPhys58_1985,PerramJW_PhysRevE54_1996,DonevA_PhysRevE75_2007}). However, without additional tests, the value of $\Lambda_1$ does not distinguish between the two antipodal ellipses represented by the same quadratic form and thus signals an overlap even when the ellipses in question are on the opposite sides of the sphere. For a usable algorithm, collisions with the antipodes of the ellipses represented by the quadratic forms must be ignored. This is handled in the following section.

\subsection{Solution branches and secondary contacts}

\begin{figure}
  \centering 
  \includegraphics[width=\linewidth]{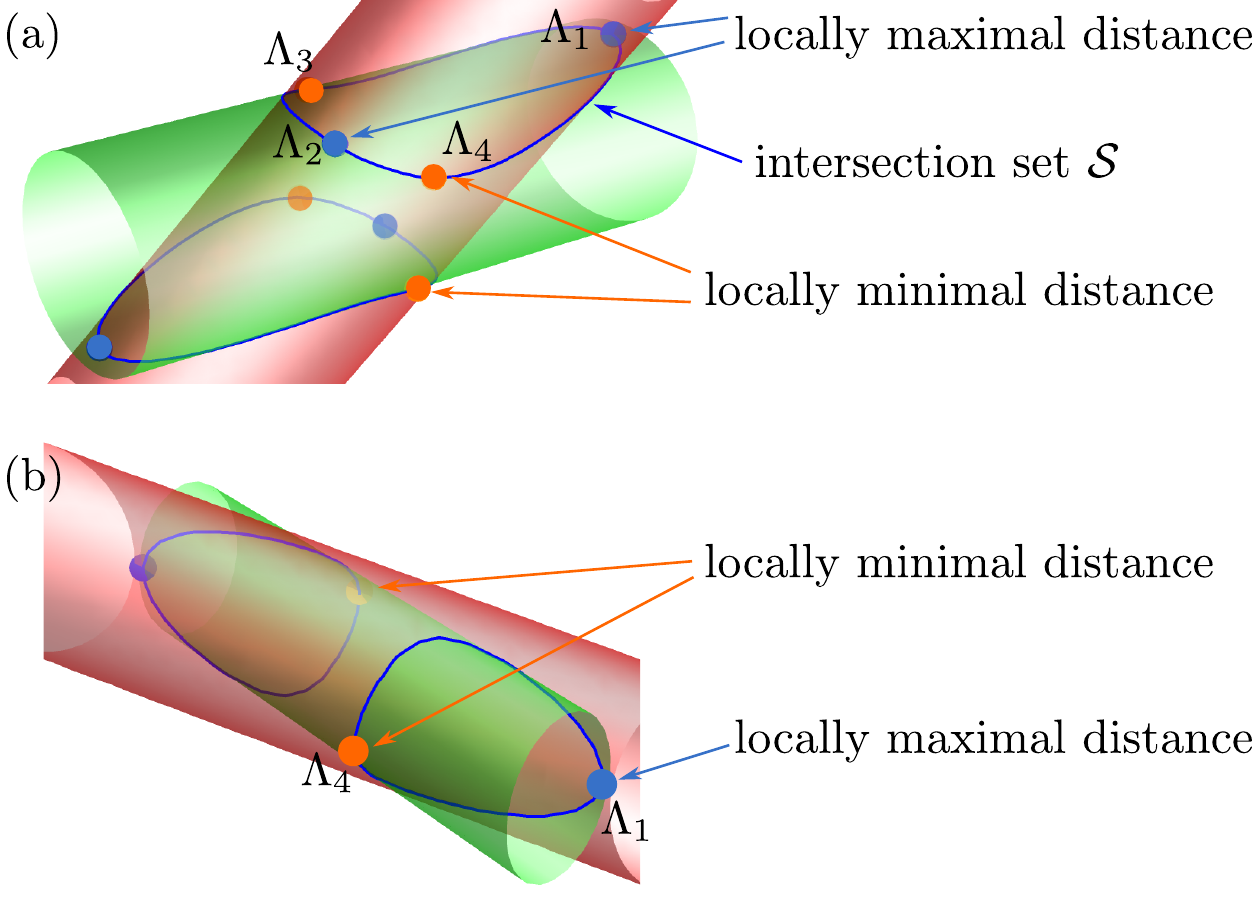}
  \caption{A generic intersection set $\mathcal{S}$ of two obliquely intersecting elliptical cylinders. The intersection consists of two antipodal loops, with two points of maximal distance and two points of minimal distance from the origin. These represent the four tangency cases; only $\Lambda_1$ and $\Lambda_2$ are relevant for our analysis. In exceptional cases, the two loops might be joined in a four-way junction.}
  \label{fig:intersect}
\end{figure}

Points of tangency of ellipses on the sphere can be defined in terms of the full intersection set of two elliptical cylinders in three dimensions,
\begin{equation}
\mathcal{S}=\{\bm{r}, a(\bm{r})=b(\bm{r})=1\}.
\end{equation}
The ellipses, obtained as intersections of $a$ and $b$ with the sphere of radius $r$, are intersecting at points on $\mathcal{S}$ at radius $r$ and are tangent in \emph{critical points} on $\mathcal{S}$ with locally extremal distance $r$ from the origin. The maximized smallest eigenvalue $\Lambda_1$, which we derived in the previous section, simply corresponds to the critical point of $\mathcal{S}$ farthest from the origin; but this is just one of the critical points. 

Degenerate cases aside, the set $\mathcal{S}$ consists of an antipodal pair of two disjoined loops. Each loop can have at most four critical points -- two with locally maximal and two with locally minimal distance to the origin, corresponding to four values $r^{-2}=\{\Lambda_1,\Lambda_2,\Lambda_3,\Lambda_4\}$ (Fig.~\ref{fig:intersect}a). Depending on the relative orientation and size of the ellipses, there may be only two critical points, $r^{-2}=\{\Lambda_1,\Lambda_4\}$ (Fig.~\ref{fig:intersect}b). At the transition between these two regimes, the critical points $\Lambda_2$ and $\Lambda_3$ merge into an inflection point before disappearing.  In other borderline cases with zero measure, the intersection set may be a ``basket'' with two four-fold junctions, or may have whole arcs at constant distance from the origin. These can all be understood as limiting cases with degenerate maxima and minima.

The maximal critical points $\Lambda_{1,2}$ correspond to the tangency with appearance of two new intersections when ellipses are stretched past the tangency condition. The minimal critical points $\Lambda_{3,4}$ correspond to the disappearance of intersections when stretching ellipses past the tangency condition. Only the maxima -- the critical points $\Lambda_{1,2}$ -- are relevant for detecting ellipse contacts. The remaining two critical points $\Lambda_{3,4}$ involve inverted ellipses, as they describe points on $\mathcal{S}$ with locally minimal distance to the origin and are thus closer than at least one of the quadratic form semiaxes.

The antipodal doubling of ellipses means that the tangency at $\Lambda_1$ may correspond to the contact with the antipode of the second ellipse, so it might not be the one we are looking for. If there are only two critical points, there is no other possible contact. If there are four critical points, growing the ellipses further makes them touch again at the next locally maximal critical point ($\Lambda_2$). This contact might be between the correct pair of ellipses, or it could be between the same pair of ellipses as the $\Lambda_1$ critical point, in which case it is not a candidate for a true contact either.

\begin{figure*}
  \centering 
  \includegraphics[width=\textwidth]{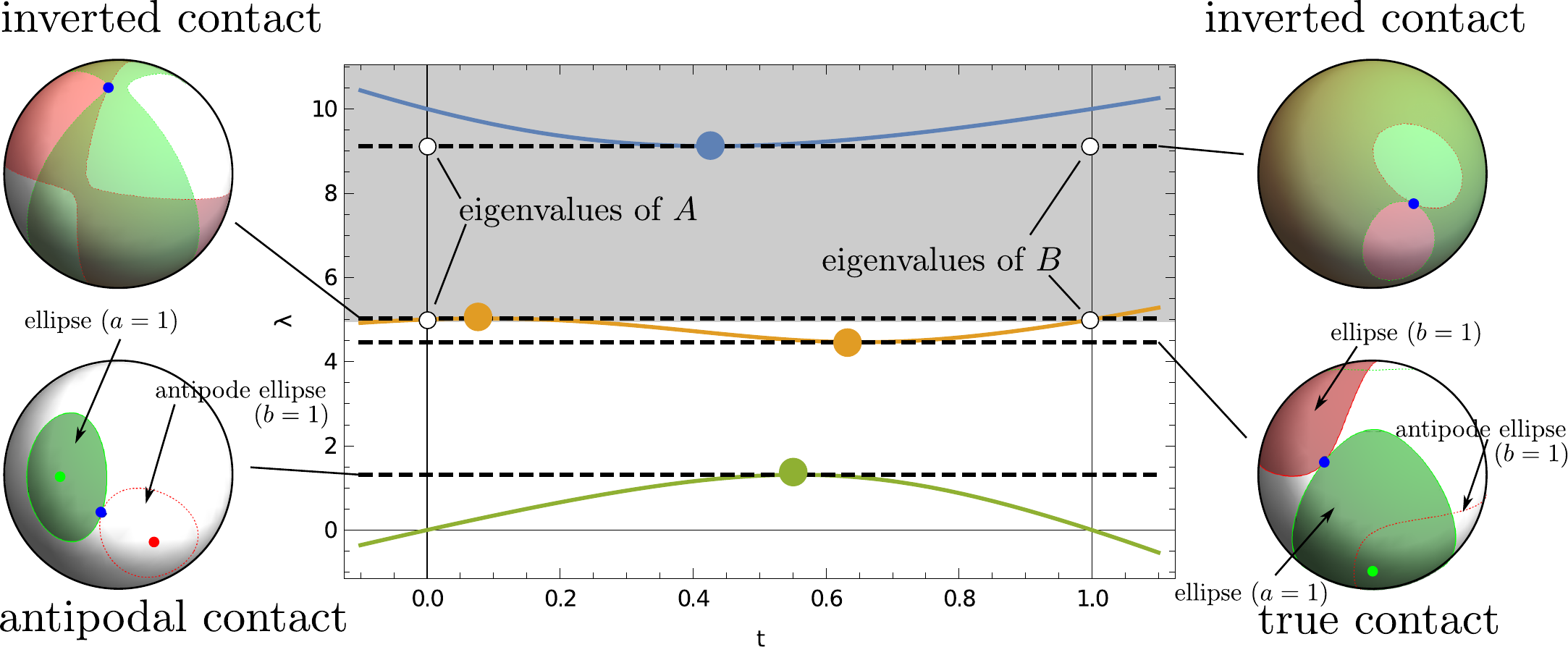}
  \caption{Eigenvalue spectrum of a generic case, with all four extrema $\Lambda_i$ occuring inside the interval $t\in(0,1)$. If the first contact is between the antipodes (lower left inset), the true contact and thus the correct value of the contact function is found by the minimum of the second eigenvalue. Observe that the green ellipse still intersects the red antipode ellipse, which we are ignoring. If the first contact is between the correct ellipses, then the lowest eigenvalue is the correct solution -- we need information about the correct antipode to test for that. The upper two extrema correspond to inverted contacts (shaded area lies above the lowest nonzero eigenvalue of $A$ and $B$).}
  \label{fig:normal}
\end{figure*}

As already discussed, the maximum of the lowest eigenvalue, $\Lambda_1$, solves for the first contact. The rest of the contacts can also be tied to extrema of the eigenvalues of $Q(t)$ over $t$. The values $\Lambda_2$ and $\Lambda_3$ correspond to the minimum and the maximum of the middle eigenvalue, and $\Lambda_4$ to the minimum of the largest eigenvalue (Fig.~\ref{fig:normal}). Unlike the lowest eigenvalue of $Q(t)$, which is guaranteed to have a local maximum between $t=0$ and $t=1$, the remaining eigenvalues can have extrema outside the interval $t\in(0,1)$, or none at all. In these cases, constrained minimization returns one of the edge points of the interval.

\begin{figure}
  \centering 
  \includegraphics[width=\linewidth]{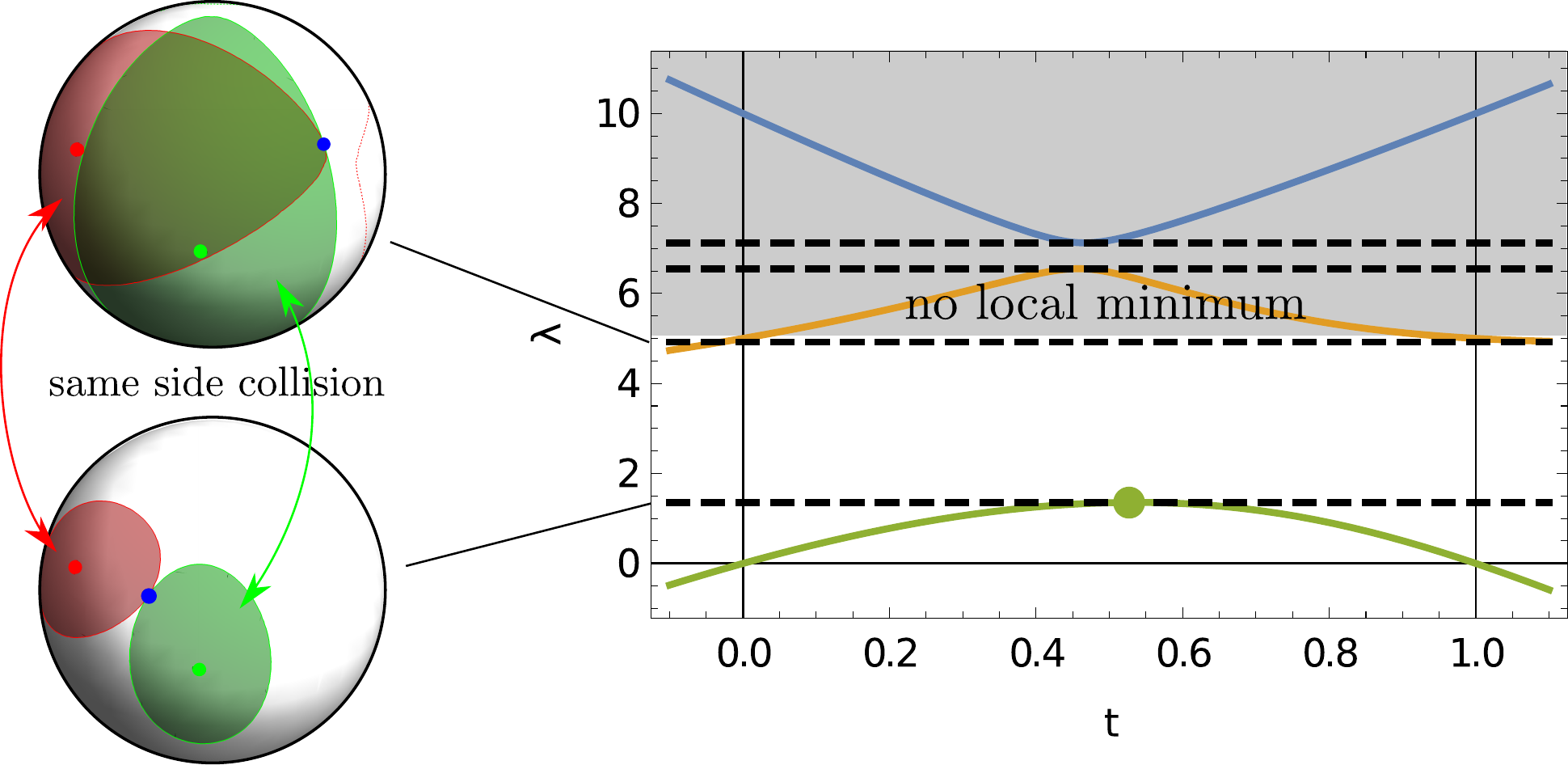}
  \caption{Eigenvalue spectrum of a case where the first two contacts are between the same ellipses, signifying that either the first contact is correct or neither of them is (as in the latter case both contacts are between antipodes). Such situations are characterized by the middle eigenvalue not having a local minimum in the interval $0<t<1$. Top two contacts are not depicted. Shaded area corresponds to inverted ellipses.}
  \label{fig:sameside}
\end{figure}

If there are only two critical points on each loop of the intersection manifold $\mathcal{S}$, then the middle eigenvalue has no local extrema, neither inside the interval $(0,1)$ nor anywhere else on the real line, and $\Lambda_{2,3}$ are undefined. If there are four critical points, the local extrema may lie outside the interval $t\in (0,1)$. This corresponds to a second contact between the same pair of ellipses as $\Lambda_1$, meaning that either both critical points signify contact between the true ellipses or both signify contact with the antipode, in which case there is no contact (Fig.~\ref{fig:sameside}). This is convenient, as simply checking for existence of a minimum of the middle eigenvalue inside the interval $t\in(0,1)$ includes all cases in which the critical point $\Lambda_2$ can constitute a real contact. Finally, if the resulting $\Lambda_1$ or $\Lambda_2$ exceed any of the eigenvalues of $A$ or $B$ (which coincide with the nonzero eigenvalues of  $Q(t=0)$ and $Q(t=1)$), it signifies a contact where at least one ellipse is inverted. We can test this by finding the minimum nonzero eigenvalue $\Omega$ of $A$ and $B$, corresponding to the largest semiaxis of the largest ellipse. Critical points that exceed this value, $\Lambda_{1,2}>\Omega$, do not correspond to valid contacts, nor can their values be unambiguously used as an analitical continuation of the contact function, because mixing of antipodes into the inverted ellipse makes the choice between the branches impossible.

\subsection{Contact function}

To define a well-behaved contact function $F$ to use as a test for ellipse-ellipse intersections, the correct eigenvalue must be selected. This is done with the help of the eigenvectors, which correspond to critical points. We denote with $\pm \bm{v}_{1}$ and $\pm \bm{v}_{2}$ the eigenvectors corresponding to $\Lambda_1$ and $\Lambda_2$, respectively, and $\bm{r}_A$ and $\bm{r}_B$ are the true centers of the ellipses on the unit sphere, their signs picking the correct ellipse of the antipodal pair. If the true ellipse collides with the antipode of the second one, the projections of the intersection vector onto the vectors of ellipse centers are of opposite signs, and vice-versa. If there is no contact, or the contact is with an inverted ellipse, assigning the value $F=\Omega$ makes the function continuous under variations of the relative position of the ellipses. The full algorithm for computing the contact function is described in Algorithm~\ref{alg:collision}. 
\begin{algorithm}[b]
\label{alg:collision}
\SetAlgoLined
\KwResult{Contact function $F$}
$\Omega\leftarrow \operatorname{min} \operatorname{eigenvalue}_{2,3}(Q(0),Q(1))$\;
$t_1 \leftarrow \operatorname{argmax}_{t\in[0,1]} \operatorname{eigenvalue}_{1}(Q(t))$\;
$\Lambda_1\leftarrow Q(t_1)$\;
$\bm{v}_1 \leftarrow \operatorname{eigenvector}(Q(t_1),\Lambda_1)$\;
\eIf{($\bf{r}_A\cdot \bm{v}_1)(\bm{r}_B\cdot \bm{v}_1)>0$}{
    \eIf{$\Lambda_1 < \Omega$}
        {\Return $\Lambda_1$\;}
        {\Return $\Omega$\;}
}
{
$t_2 \leftarrow \operatorname{argmin}_{t\in[0,1]} \operatorname{eigenvalue}_{2}(Q(t))$\;
$\Lambda_2\leftarrow Q(t_2)$\;
    \eIf{$0<t_2<1$ {\normalfont\textbf{and}} $\Lambda_2<\Omega$}{
        \Return $\Lambda_2$\;
    }
    {
        \Return $\Omega$\;
    }
}
\caption{Ellipse-ellipse contact function}
\end{algorithm}

The contact function $F$, which is according to the above criterion equal to $\Lambda_1$, $\Lambda_2$, or $\Omega$, can be used either to directly detect when ellipses overlap ($F<1$) or to construct a pair potential. Instead of a hard core repulsion, a soft repulsion potential for overlapping cases $F<1$ can be defined based on the value of $F$, such as $-\ln F$, $F^{-1}$, $F^{-1}+F-2$, or $1-F$, the last being a soft potential of finite strength at complete overlap. On the other hand, long-range values of $F>1$ could act as a distance metric, e.g., in a Lennard-Jones-like potential, as they do in Euclidean space~\cite{EveraersR_PhysRevE67_2003}. Setting the function to $\Omega$ in  cases for which the ellipses cannot intersect no matter the stretch factor, ensures a constant potential and zero force on the particles for that entire region, and makes the function well-behaved for use in methods that require a potential (e.g., Monte Carlo methods). Even though there is no correspondence between such an artificially fashioned potential and any physical phenomena we know of, such an academic exercise could provide a reasonable approximation to medium-range behavior that could match empirical observations in certain physical systems.

\begin{figure}
  \centering 
  \includegraphics[width=\linewidth]{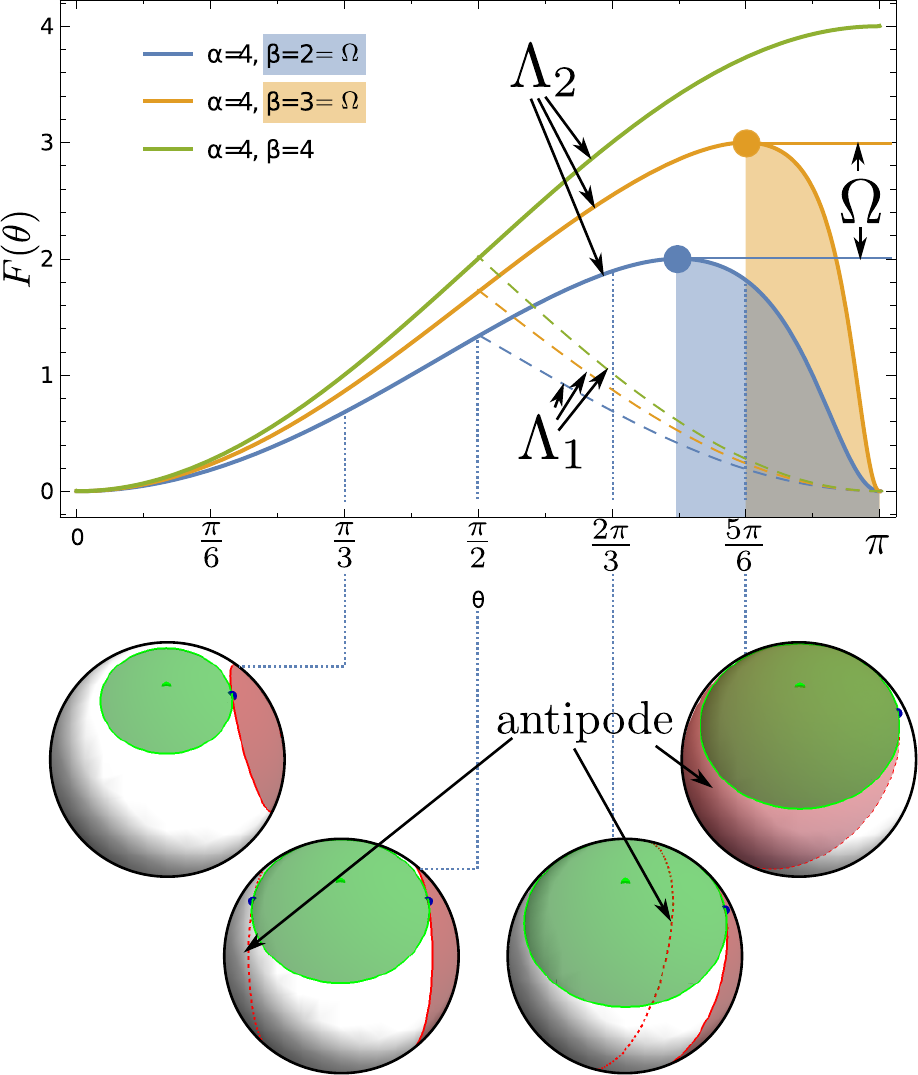}
  \caption{Contact function for circles of different relative radii separated by angle $\theta$. Transition to the antipodal contact at $\theta>\tfrac{\pi}{2}$ is continuous; the insets show that the secondary contact is the correct one, while the contact with the antipode (dashed circle outline without infill) is ignored. If the circles are of the same size, the contact function is monotonously increasing, but if they are of different sizes, the decreasing part (shaded below the curve) corresponds to the case when the ellipses cannot be made to touch by stretching, and the corresponding eigenvalue detects the second contact between the ``wrong'' pair of ellipses. In this region, the value of $F$ is set to $\Omega$ (horizontal line), which corresponds to the inverse square radius of the larger circle. Parameters $\alpha$ and $\beta$ correspond to inverse square radii (see Eq.~\eqref{eq:alphabeta}).
  }
  \label{fig:circles}
\end{figure}

\section{Examples}

\subsection{Intersection of unequal circles}

The simplest example that can be used for interpretation of the contact function $F$ is a pair of unequal circles. Define the following pair of quadratic forms:
\begin{eqnarray}
\nonumber a&=&\alpha(x^2+y^2),\\
b&=&\beta(x^2\cos^2 \theta+z^2\sin^2\theta-2xz\cos\theta\sin\theta+y^2),
\label{eq:alphabeta}
\end{eqnarray}
with $\alpha>\beta>1$ and $\theta$ the angular separation of the circle centers. In this case, the extremal eigenvalues (without applying the restriction to $0<t<1$) have a relatively simple closed form, and the contact function can be expressed as
\begin{equation}
  F(\theta)=\frac{\alpha\beta\sin^2\theta}{\alpha+\beta+2\sqrt{\alpha\beta}\cos\theta}
\end{equation}
The function's behavior with respect to $\theta$ is depicted in Fig.~\ref{fig:circles} for a few combinations of circle sizes $\alpha$ and $\beta$. At $\theta>\tfrac{\pi}{2}$, we have $F(\theta)=\Lambda_2$, corresponding to the second contact, as the first contact is with the antipode. We observe that the crossover between the branches is continuously differentiable. However, with the exception of equal circles, we see that the function reaches a maximum at $\cos^2\theta=\frac{\beta}{\alpha}$ and then goes back to zero at $\theta=\pi$. This part of the contact function corresponds to the second collision also being with the antipode. The collision is internal (non-facing normals), and the interpolation parameter at minimal middle eigenvalue is $t>1$. In our algorithm, we assign these collisions $F=\Omega$.

\subsection{Computational cost of the algorithm}

\begin{figure}
  \centering 
  \includegraphics[width=\linewidth]{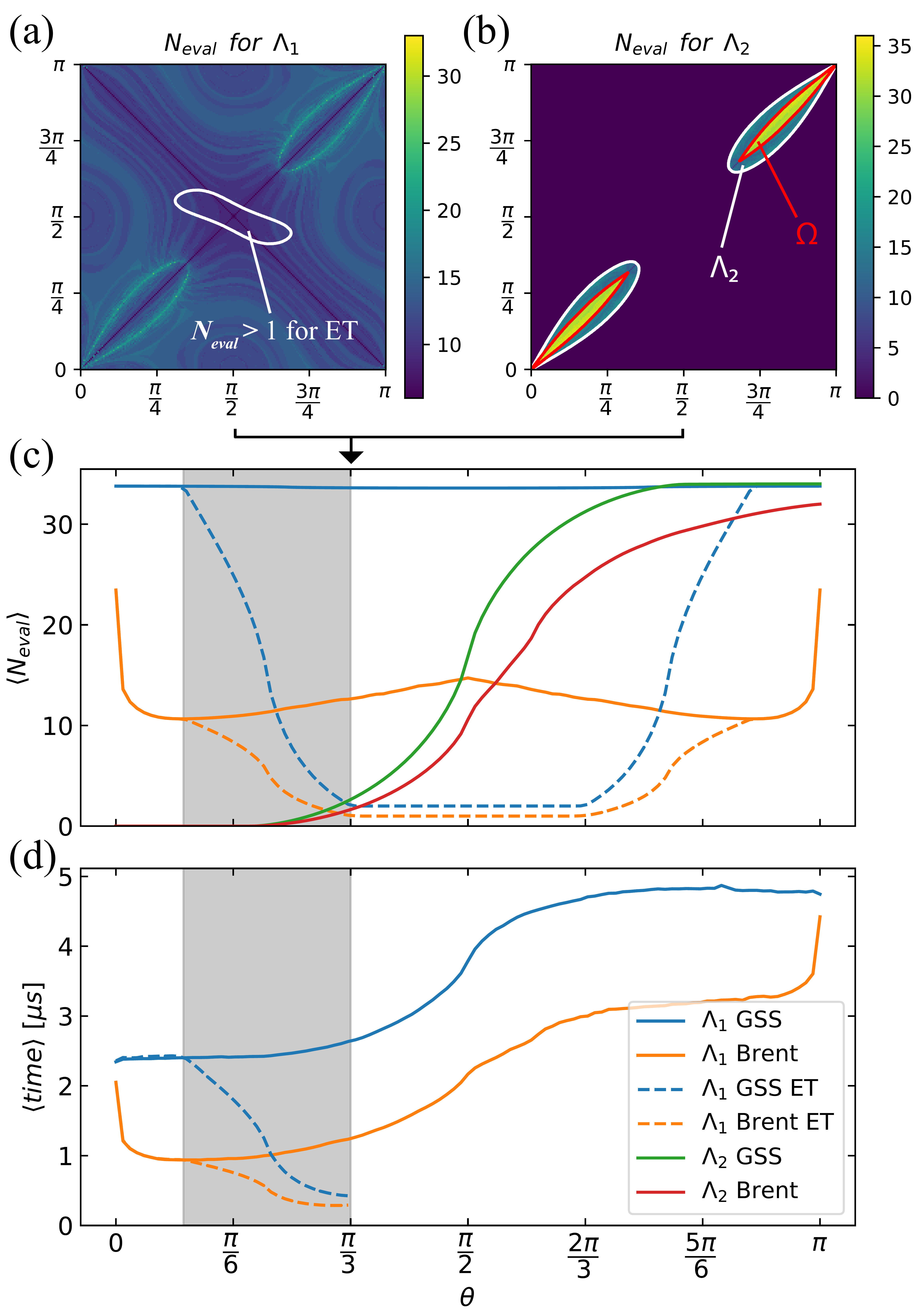}
\caption{Number of eigenvalue evaluations and contact function calculation times for a pair of ellipses with $\xi_1=0.5$ and $\xi_2=0.15$. Number of evaluations needed to determine \textbf{(a)} $\Lambda_1$ and \textbf{(b)} $\Lambda_2$ for the Brent method at angular separation $\theta=\pi/3$ in the whole orientational domain. For $\Lambda_2$, $N_{eval}=0$ in a large part of the domain where $\Lambda_1$ is the correct eigenvalue for determining the contact function (outside of the white contours in panel (b)). Around this eigenvalue crossover, the number of $\Lambda_1$ evaluations is increased. The white contour line in panel (a) surrounds the region where $N_{eval}>1$ even with the ET enabled. The red contours in panel (b) show the border where the contact function transitions to the constant value of $F=\Omega$. \textbf{(c)} Number of contact evaluations for GSS and Brent line minimizators. The increase in $\Lambda_2$ evaluations is a consequence of growing regions where $\Lambda_1$ is not the correct eigenvalue. ET (dashed lines) significantly decreases the number of $\Lambda_1$ evaluations. The gray area corresponds to distances where the overlap of ellipses depends on their orientations; on the left side of this region, overlap is guaranteed, while there can be no overlap on the right side of the region. \textbf{(d)} Comparison of contact function calculation times for GSS and Brent methods. ET results are relevant only for $\theta<2\arcsin{\xi_1}$, as they can only be used to determine overlap/no overlap.} 
  \label{fig:evaluations}
\end{figure}

The algorithm itself is fast, as the eigenvalue calculation can be expressed in a closed form, although it uses trigonometric functions which are slower than simple multiplications. One-dimensional minimization and maximization routines are available in any number of numerical libraries. We implemented two such routines, the golden section search (GSS) and the Brent method (GSS with quadratic interpolation), and compared both the numbers of eigenvalue evaluations $N_{eval}$ to achieve the desired accuracy (tolerance of $10^{-7}$ in $t$) as well as calculation times. In Fig.~\ref{fig:evaluations}, we show the results for a pair of ellipses on a unit sphere with major and minor semiaxes $\xi_1=0.5$ and $\xi_2=0.15$, respectively (aspect ration $\varepsilon=\xi_1/\xi_2\approx3.33$). At a given angular separation $\theta$, the contact function and the computational cost required to determine it with the Brent method depend on orientations of both ellipses as shown for $\theta=\pi/3$ in Figs.~\ref{fig:evaluations}a and \ref{fig:evaluations}b for the number of first and second eigenvalue evaluations, respectively. The number of evaluations for $\Lambda_1$ mostly lies between $10$ and $20$, with the exceptions of diagonals with fewer evaluations and two loops with $N_{eval}\sim 30$ that correspond to the cases near the $\Lambda_1$ and $\Lambda_2$ crossover. For relatively high aspect ratios $\varepsilon$, as is the case in the demonstrated example, the second derivative close to the crossover becomes large, which is unfavorable for the Brent minimization. Inside these loops, the second eigenvalue becomes relevant for the contact function, as indicated in Fig.~\ref{fig:evaluations}b ($\Lambda_2$ is only evaluated in regions where the $\Lambda_1$ eigenvector test fails, see Algorithm~\ref{alg:collision}). Additionally, closer to the diagonals, the local minimum of $\lambda_2(t)$ inside the interval $[0, 1]$ disappears and the algorithm returns $\Omega$. Note again that despite the algorithm branch changes, the contact function is continuous in the whole configuration space. 

We evaluate the necessary computational cost to determine the contact function both for the GSS and Brent methods. The results with respect to the angular separation $\theta$ are shown in Fig.~\ref{fig:evaluations}c, where the value at each distance represents the average number of eigenvalue evaluations over the whole orientational domain ($10^6$ points, Figs.~\ref{fig:evaluations}a and \ref{fig:evaluations}b). The number of $\Lambda_1$ evaluations with the GSS method remains (almost) constant for all distances, as a fixed number of interval divisions is necessary to achieve the desired precision. This number is also markedly higher compared to the Brent method, which shows that quadratic interpolation is highly effective for this problem (this could be expected from eigenvalue curves in Figs.~\ref{fig:normal} and \ref{fig:sameside}). Note that the number of $\Lambda_1$ evaluations is symmetric around $\theta=\pi/2$, as elliptical cylinder configurations are invariant to coordinate transformation $\theta\rightarrow\pi-\theta$ and only the antipode interpretations for the correct/wrong ellipse are exchanged. The number of $\Lambda_2$ evaluations does not show this symmetry. At small angular separations, the first eigenvalue will always be the correct one and only for higher $\theta$ does the second eigenvalue evaluation become necessary in parts of the orientational space (Fig.~\ref{fig:evaluations}b). These regions become larger as $\theta$ is increased (at some point, they consume the whole orientational domain), which in turn increases the average $\Lambda_2$ evaluation numbers.

In some situations, e.g., for simulations of hard particles, the calculation of the exact contact function is not needed. The optimization algorithm can be terminated immediately after a value of $\Lambda_1>1$ is encountered, as that means no overlap. The average number of $\Lambda_1$ evaluations with this early termination (ET) condition is shown in Fig.~\ref{fig:evaluations}c with dashed lines and leads to a sharp decrease of the necessary calculations in a large part of the plot. As shown in panel (a) for $\theta=\pi/3$, more than one $\Lambda_1$ evaluation is necessary only inside the white contour which grows/shrinks for smaller/larger distances. Additionally, the gray region in the plot highlights the distances where the overlap appears only for certain ellipse orientations ($2\arcsin{\xi_2}\leq \theta\leq 2\arcsin{\xi_1}$) -- on the left side of this region, ellipses overlap for all orientations and on the right, overlap is not possible as they are too distant and the eigenvalue calculation can be skipped entirely.

Finally, Fig.~\ref{fig:evaluations}d shows the average calculation time to evaluate the contact function. The results are on the order of $\si{\micro s}$ and closely follow the combined number of $\Lambda_1$ and $\Lambda_2$ evaluations from panel (c), with the increase in calculation times corresponding to additional evaluations needed to determine the second eigenvalue at larger distances. If ET is enabled, the efficiency of the calculation is significantly improved.

\subsection{Dense packings of spherical ellipses}

\begin{figure}
  \centering 
  \includegraphics[width=\linewidth]{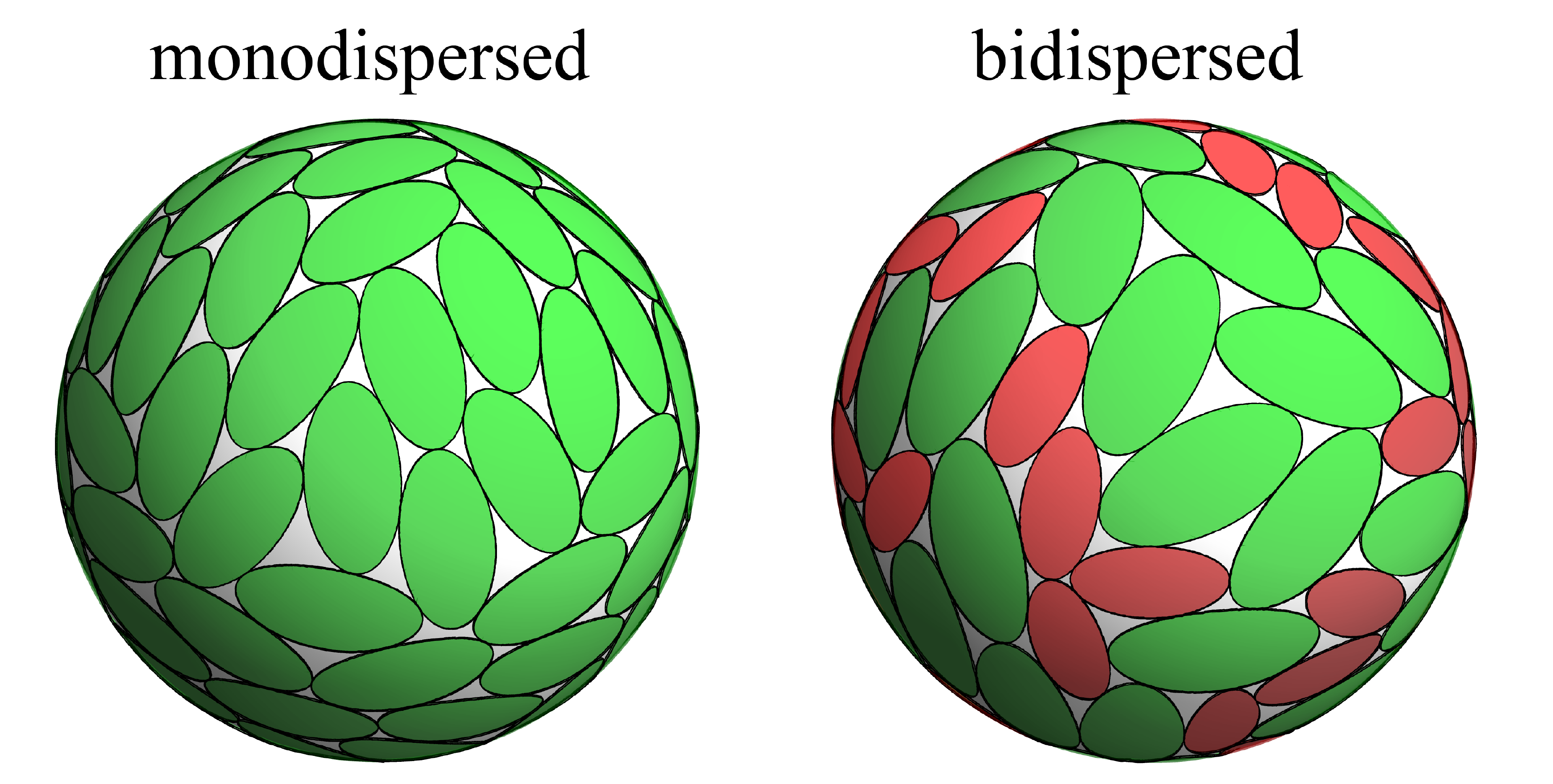}
\caption{Examples of monodispersed and bidispersed dense packings of $N=100$ spherical ellipses with aspect ratio $\varepsilon=2$. In the bidispersed case, half of all ellipses are smaller by a factor of $1.4$.}
  \label{fig:mono_bi_example}
\end{figure}

To demonstrate the use case of the proposed algorithm in multiparticle simulations, we calculated dense packings of $N=100$ spherical ellipses with $\varepsilon=2$ for both monodispered and bidispersed systems (Fig.~\ref{fig:mono_bi_example}). We employed an energy minimization-based approach similar to the scheme used by \citet{Mailman2009PRL} where the system is randomly initialized at a packing fraction far from the jamming point, with subsequent iterative increases of particle sizes and relaxations to remove all overlaps. As angular separation $\theta$ between the centers of neighboring (touching) ellipses is smaller than $\pi/2$ for our system parameters ($N$ and $\varepsilon$), it is sufficient to calculate only the minimal first eigenvalue to determine the contacts -- possible cases with antipodal contacts can be excluded based on ellipse separation alone.  

\section{Discussion}

Depending on the requirements, the algorithm can be optimized further. For example, with SIMD instructions, evaluations at multiple $t\in (0,1)$ could be performed with minimal overhead, allowing for faster determination of the correct eigenvalue branch and narrower initial bracket for the optimization algorithm. For the purposes of collision-driven molecular dynamics, the expensive $\mathcal{O}(n^2)$ complexity of evaluating pair interactions for a large number of particles can be alleviated by keeping track of nearest neighbours (e.g., by adapting pre-existing methods that make use of the contact functions \cite{DonevA_JComputPhys202_2005}). Tracking and changing particle positions and orientations while keeping their shapes constant requires keeping track of the rotation matrices in a numerically stable form, which can be done either by tracking the ellipse center vectors and the vector of its principal component (e.g., through Euler angles) or by using unit quaternions.

Our algorithm is largely based on the algorithm of \citet{PerramJW_JComputPhys58_1985} but has some important differences due to the differences between spherical and Euclidean geometry. On the one hand, spherical geometry of the problem makes it simpler, because in the Euclidean space, translations and rotations have to be considered separately, while on the sphere, the only parameter for the position and orientation of the ellipse is a single rotation matrix. Similarities can be partially restored by handling the Euclidean case in homogeneous (projective) coordinates, but then the confinement surface is a plane, not a sphere, resulting in a different algorithm. Due to this difference, our algorithm requires solving an eigenvalue problem and not a linear system of equations. In general, the eigenvalue problem is numerically more expensive, but for $3\times 3$ matrices, a closed-form solution is available.

From the aspect of finding the correct solutions, the spherical version of the algorithm is more involved, as the configuration space of possible intersections is topologically nontrivial and splits into different parts based on the behavior of eigenvalue bands with respect to the parameter $t$. The antipodal doubling means we need additional information to treat different branches of the solution differently. However, as shown in our work, this can be done with a few trivial tests, with the only caveat that the long-range contact function is spliced and undefined (clipped to $\Omega$) in parts of the configuration space.

\section{Conclusions}

The simplicity and speed of the presented algorithm makes it a viable workhorse for future simulations on a sphere, be it interactions of hard particles or general long-range interactions where distances are needed, although the concept of the contact function as a distance metric must be considered with care. Collision detection and generalized distance can be used for Monte Carlo simulations, while molecular dynamics can make use of the intersection vector and the normal vector to the surface as well. Elongation of particles is known to affect optimal packing fraction of random packings in Euclidean space~\cite{DelaneyG_PhilosophicalMagazineLetters85_2005,DonevA_Science303_2004}, and with the presented algorithm, related questions can be answered for packings on a sphere.

Simulations can also be augmented with other potentials that do not utilize the contact function -- for example, multipolar interactions, which may account for elliptical magnetic particles or electrostatically charged macromolecules. The algorithm is viable for particles of different aspect ratios and sizes, so it can be used for simulations of polydisperse particle systems. Another important use case is in representation of arbitrarily shaped objects as isosurfaces of Gaussian sums (called blobs or metaballs in 3D graphics). A product of Gaussians, resulting directly in addition of quadratic forms when constraned to a sphere, also resembles posterior Bayesian update when handling probability models for directional or geographical data, which may be relevant in data processing and machine learning.

Finally, more fundamental questions can also be tackled. Recall that both the Tammes problem and its long-range potential cousin, the Thomson problem, have been well-studied not only by physicists, but also from the perspective of fundamental and applied mathematics and computer science. Generalization to an anisotropic case is a richer example, which without doubt hides many undiscovered facts about spherical packings.

\section*{Acknowledgments}
We acknowledge support by Slovenian Research Agency (ARRS) under Contracts No.\ P1-0099 and No.\ J1-9149. The work is associated with the COST Action No.\ CA17139.

\bibliographystyle{apsrev4-1}
\bibliography{bibliography}

\end{document}